\documentclass[soumission]{jsfds}
\usepackage[latin1]{inputenc}
\usepackage[T1]{fontenc}

\usepackage{amsfonts, amsmath, amssymb}
\usepackage{dtklogos}
\usepackage{url}
\usepackage{natbib}
\usepackage{float}
\usepackage{rotating}
\usepackage[]{algorithm2e}
\usepackage{ulem}
\startlocaldefs

\def\1{1\kern-.20em {\rm l}}
\endlocaldefs

\setmainlanguageenglish
\firstpage{1} 

\newtheorem{prop}[theo]{Property}
\newtheorem{propo}[theo]{Propostion}

\newcommand{\modif}[2]{{}{#2}}
\begin{document}

\begin{frontmatter}


  \titre{A pruned dynamic programming algorithm to recover the best segmentations with $1$ to $K_{max}$ change-points.}
  \titrecourant{pDPA to recover the best segmentations with $1$ to $K_{max}$ change-points.}

  \begin{InfosAuteurs}
    \auteur{%
      \prenom{Guillem} 
      \nom{Rigaill}%
      \thanksref{t1}%
      \contact[label=e1]{rigaill@inra.evry.fr}%
    }%

    \affiliation[t1]{Institut of Plant Sciences Paris-Saclay (IPS2), UMR 9213/UMR1403, CNRS, INRA, Université Paris-Sud, Université d'Evry, Université Paris-Diderot, Sorbonne Paris-Cité, Bâtiment 630, 91405 0rsay, France\\
      \printcontact{e1}%
    }

    \enteteauteurs{Rigaill} 
  \end{InfosAuteurs}

  \begin{abstract}
A common computational problem in multiple change-point models is 
to recover the segmentations with $1$ to $K_{max}$ change-points of minimal cost with respect to some loss function.
Here we present an algorithm to prune the set of candidate change-points which is based on 
a functional representation 
of the cost of segmentations. 
We study the worst case complexity of the algorithm when there is a unidimensional
parameter per segment and demonstrate that it is at worst equivalent to 
the complexity of 
the segment neighbourhood algorithm: 
$\mathcal{O}(K_{max} n^2)$.
For a particular loss function we demonstrate that pruning is on average efficient even 
if there are no change-points in the signal.
Finally, we empirically study the performance of the algorithm in the case of the quadratic loss and show that it is faster than the segment neighbourhood algorithm.
\end{abstract}

  \begin{motscles}
    \mot{multiple-change-point detection}
    \mot{dynamic programming}
    \mot{functional cost}
    \mot{segment neighbourhood}
  \end{motscles}

 
\end{frontmatter}

\newcommand{\Mc}{\mathcal{M}}
\newcommand{\Oc}{\mathcal{O}}
\newcommand{\Sc}{\mathcal{S}}
\newcommand{\Tauc}{\mathcal{\tau}}
\newcommand{\Cc}{C\hspace{+.1ex}}
\newcommand{\FCc}{\widetilde{C}\hspace{+.1ex}}
\newcommand{\SumSegmentsZeroK}{\sum_{k=1}^{K+1}}
\newcommand{\SumSegmentsZeroKminusOne}{\sum_{k=1}^K}
\newcommand{\SumSegmentkDataPoint}{\sum_{t=\tau_{k-1}}^{\tau_{k}-1}}
\newcommand{\SumSegmentKDataPoint}{\sum_{t=\tau_{K}}^{\tau_{K+1}-1}}

\newcommand{\SumSegmentTautoT}{\sum_{i=\tau}^{t-1}}
\newcommand{\SumSegmentTautoTplusOne}{\sum_{i=\tau}^{t}}

\newcommand{\CostSegmentK}{c_{\tau_{K}:\tau_{K+1}}}
\newcommand{\CostSegmentk}{c_{\tau_{k-1}: \tau_{k}}}
\newcommand{\CostSegmentTauT}{c_{\tau \hspace{-.0ex}:\hspace{+.2ex} t}}
\newcommand{\CostSegmentTauTprime}{c_{\tau \hspace{-.0ex}:\hspace{+.2ex} t'}}
\newcommand{\CostSegmentTauprimeT}{c_{\tau' \hspace{-.1ex}:\hspace{+.2ex} t}}
\newcommand{\CostSegmentTauprimeTprime}{c_{\tau' \hspace{-.1ex}:\hspace{+.2ex} t'}}

\newcommand{\FCostSegmentK}{\widetilde{c}_{\hspace{+.1ex} \tau_{K}:\tau_{K+1}}}
\newcommand{\FCostSegmentk}{\widetilde{c}_{\hspace{+.1ex}\tau_{k-1}: \tau_{k}}}
\newcommand{\FCostSegmentTauT}{\widetilde{c}_{\hspace{+.1ex}\tau \hspace{-.0ex}:\hspace{+.2ex} t}}
\newcommand{\FCostSegmentTauTprime}{\widetilde{c}_{\hspace{+.1ex}\tau \hspace{-.0ex}:\hspace{+.2ex} t'}}
\newcommand{\FCostSegmentTauprimeT}{\widetilde{c}_{\hspace{+.1ex}\tau' \hspace{-.1ex}:\hspace{+.2ex} t}}
\newcommand{\FCostSegmentTauprimeTprime}{\widetilde{c}_{\hspace{+.1ex}\tau' \hspace{-.1ex}:\hspace{+.2ex} t'}}

\newcommand{\Gammai}{\gamma(Y_i, \mu)}
\newcommand{\Gammat}{\gamma(Y_t, \mu)}
\newcommand{\Kmax}{K_{max}}

\newcommand{\OnetoT}{\small{1:\hspace{+.1ex}t}}
\newcommand{\OnetoTau}{\small{1\hspace{-.0ex}:\tau}}
\newcommand{\OnetoN}{\small{1:\hspace{+.1ex}n}}
\newcommand{\ItoJ}{\small{i\hspace{-.0ex}:j}}

\newcommand{\argmin}{\mathop{\mathrm{arg\ min}}}
\newcommand{\Cost}{\mbox{\bf{Cost}}\hspace{+.1ex}}
\newcommand{\FCost}{\widetilde{\mbox{\bf{Cost}}\hspace{+.1ex}}}
\newcommand{\OPCost}{\mbox{\bf{OPCost}}}
\newcommand{\llbracket}{[\hspace{-.05ex}[}
\newcommand{\rrbracket}{]\hspace{-.05ex}]}
\newcommand{\ConvBunch}{\mathcal{B}}
\newcommand{\ConcBunch}{\mathcal{B}}
\newcommand{\rank}{\mathcal{R}}
\newcommand{\order}{O}
\newcommand{\UnN}{\{ 1 \ldots n \}}
\newcommand{\UnK}{\{ 1 \ldots k \}}
\newcommand{\ZeroN}{\llbracket 0, n \rrbracket}
\newcommand{\UnNPrime}{\llbracket 1, n' \rrbracket}
\newcommand{\UnNmoins}{\llbracket 1, n-1 \rrbracket}
\newcommand{\UnNplus}{\llbracket 1, n+1 \rrbracket}

\section{Introduction}

A common computational problem in multiple change-point models is 
to recover segmentations with $1$ to $\Kmax$ change-points of minimal cost where the cost is some well chosen
criteria, such as minus the log-likelihood or the quadratic loss
\citep{BP03,picard2005statistical,harchaoui2007retrospective,guedon2007analyzing,killick2011changepoint,
arlot2012kernel,cleynen2013segmentation,cleynen2014comparing}.
Many algorithms have been proposed to exactly solve this problem \citep{bellman1961approximation,fisher1958grouping,auger1989algorithms,
BP03,guedon2008exploring}. 
All these are dynamic programming algorithms 
and have a complexity 
which is linear in $\Kmax$ and quadratic in the length of the signal $n$: $\Oc(\Kmax n^2)$.
We will refer to these algorithms as segment neighbourhood algorithms.

In practice, this quadratic complexity in $n$ is \modif{}{a problem.}
Indeed, if $n$ is larger than $10^5$ or $10^6$ one run of the algorithm can take several hours or even days.
In applications such as DNA copy number studies the signal length is typically of this order: $n = 10^5$-$10^6$.
Several strategies have been developed to cope
with this problem, the most famous is probably the binary segmentation heuristic 
\citep{scott1974cluster}.
A common idea is to first identify a restricted set of candidate change-points 
with any fast heuristic and then
run a segment neighbourhood algorithm on this restricted set \citep{kolesnikov2003reduced,harchaoui2010multiple,gey2008using}.
These heuristics typically have a complexity which is linear in $n$.
From a computational perspective the main drawback of these approaches is that they lack optimality.

Most algorithms and heuristics that aim at recovering the 
segmentations of minimal cost 
operate on segmentations through their costs.
Here we propose a new representation of the segmentations that we call the functional cost 
where the cost of a segmentation is represented as a function of a, possibly multidimensional, parameter.
We demonstrate that using this functional cost it is theoretically possible to prune the set of segmentations 
while searching for segmentations of minimal costs with $1$ to $\Kmax$ change-points
and thus to carry out the calculation
only for a small subset of candidate segmentations. 
We call the resulting algorithm pruned dynamic programming algorithm (pDPA).

\paragraph{}
From an intuitive point view, if we consider a random sequence of $n$ data-points with $K$ true abrupt change-points, 
we expect that the segmentation with those $K$ change-points will  greatly outperform 
other possible \modif{}{segmentations} in terms of cost. Hence it makes sense that it is possible to efficiently prune the set of segmentations.
On the contrary, if we now consider a random sequence of $n$ data-points without any change-points, 
all segmentations 
will
have roughly the same cost and we do not expect an efficient pruning of the set of segmentations.
This intuition is indeed true if segmentations are represented through their cost.
In the context of another change-point optimization problem, the optimal partitioning problem \citep{jackson2005algorithm}, 
this idea was made particularly clear by \citet{killick2012optimal}. 

\modif{}{For this problem,} the PELT algorithm\modif{}{, that prunes segmentations based on their cost}, was proven to be linear if the true number of change-points is linear in the number of data-points. 
If it is not linear\modif{}{, in particular if there are no change-points, } PELT's pruning is less efficient.

Here we argue \modif{}{that if we consider the functional costs of segmentations, rather than their costs, we can have an efficient pruning even for random sequences without any change-points.}
As a proof of concept we study the complexity of the pDPA for change-point models with a unidimensional parameter per segment.
In that case, we demonstrate that the algorithm is at worst as efficient as segment neighbourhood algorithms.
For a special loss function we demonstrate that on average pruning is efficient
even if we consider random sequences without change-points and we retrieve an average complexity which is sub-quadratic in $\Oc(n\log(n))$. \modif{}{W}e implemented the algorithm for the commonly used quadratic loss and empirically show that it is faster than segment neighbourhood algorithms.

\paragraph{Related works}
Since the first pre-print of this work \modif{}{\citep{rigaill2010pruned}}, the pDPA has been implemented for other losses \citep{cleynen2014segmentor3isback},
and \modif{}{in the context of DNA copy number analysis its competitive runtime compared to the segment neighbourhood algorithm, PELT or other approaches for this problem} was confirmed by others \citep{hocking2013learning,hockingspec,maidstone2014optimal}.
Furthermore, we demonstrate in this new version of the paper, for
a simple, yet non-trivial, loss function that on average the pDPA pruning is efficient even if we consider random sequences without change-points.

Finally, the main contribution of this paper is of a computational nature, {\it{i.e}} an algorithm that recovers the segmentations with $1$ to $\Kmax$ change-points of minimal cost. 
From a computational point of view several methods boil down to this specific problem.
Importantly, the statistical properties of these methods have been studied theoretically and empirically through 
simulations and on real data by many 
\citep{yao1989least,horvath1993maximum,lavielle2000least,lavielle2005using,lebarbier2005detecting,birge2007minimal,boysen2009consistencies,
cleynen2013segmentation,zhang2007modified,
lai2005comparative,cleynen2014comparing},
and recently the pDPA as implemented in the R cghseg package for the quadratic loss 
was shown to reach state of the art performances for the segmentation of DNA copy number profiles 
\citep{hocking2013learning}.

\paragraph{Outline}
Section \ref{section:framework} \modif{}{describes} the change-point framework used in the paper.
Section \ref{section:pre_algo} \modif{}{gives} a quick overview of segment neighbourhood algorithms 
and \modif{}{informally describes} the functional cost and the pDPA.
Section \ref{section:algo} gives a detailed description of the functional cost and the pDPA. 
\modif{}{S}ection \ref{section:complexity} is about the worst case, average and empirical complexity of the
pDPA.

\section{Change-point framework and cost minimization}\label{section:framework}


\paragraph{Data}
We assume that we have a sequence of $n$ observations. 
We denote this sequence $Y = \{Y_t\}_{1 \leq t \leq n}$ and denote 
$Y_{\ItoJ}$ a 
subsequence
between data-point $i$ and data-point $j-1$, 
{\it{i.e.}} $Y_{\ItoJ} = \{Y_t\}_{i \leq t < j}.$


\paragraph{Segmentations}
We define a segmentation $m$ of $Y$ by a set of $K$ change-points 
splitting the sequence in $K+1$ segments.
We denote the positions of these $K$ change-points $\tau_k$ for $k=1, \ldots, K$ and
we also set the convention that $\tau_0=1$ and $\tau_{K+1}=n+1$.
We call $r_k$ the $k$-th segment of a segmentation: 
$r_k= \tau_{k-1}:\tau_{k} =\{ i | \tau_{k-1}  \leq i < \tau_{k} \}$.
$\tau_{k-1}$ is thus the first data-point belonging to $r_k$ 
and $\tau_{k}$ is the first data-point not belonging to $r_k$.
\modif{}{F}or $K > 0$, we define $\Mc_{\OnetoT}^K$ as the set of 
all possible segmentations with exactly $K$ change-points 
of the sequence $Y_{\OnetoT}.$
There are ${t-2\choose K}$ such segmentations 
\modif{}{and so for the whole sequence we have $|\Mc_{\OnetoN+1}^K| = {n-1\choose K}.$} 

\paragraph{Statistical model}
Our goal is to infer from the data both the positions and the number of change-points.
Many methods, statistical models and model selection criteria have been proposed to address this problem
for various types of data and various types of changes,
{\it{i.e.}} change in the mean, in the variance, in the distribution etc$\ldots$\modif{}{\citep{lavielle2005using,cleynen2013segmentation,arlot2012kernel}}.
In practice most of these methods critically rely on algorithms or heuristics that explore
the set of all possible segmentations in search for a list of optimal segmentations w.r.t.
some well chosen statistical criteria. However, this set of segmentations is extremely large 
($|\Mc_{\OnetoN+1}^K| = {n-1\choose K}$) and it is hard 
to explore it efficiently, especially when $n$ is large.

From a computational point of view this exploration problem is often re-formulated 
as a cost minimization problem under the constraint that the number of change-points is 
$K=1, 2 \cdots \Kmax$, where $\Kmax$ is defined by the user 
\citep{bellman1961approximation,fisher1958grouping,auger1989algorithms,
hawkins2001fitting,BP03,harchaoui2007retrospective,guedon2008exploring,arlot2012kernel,
cleynen2013segmentation,guedon_hal_00850847}. 
To be more specific, the aim is to find the segmentations of minimal cost in $\Mc_{\OnetoN+1}^K$ for $K$ 
from $1$ to $\Kmax$, where the cost of a segmentation
is the sum of the costs of its segments:
\begin{equation}\label{eq:SegAddCost}
R_m = \SumSegmentsZeroK \CostSegmentk,
\end{equation}
with $\CostSegmentk$ is the cost of the $k$-th segment of $m$. 

\paragraph{A smaller class of problems}
In this paper, we consider a smaller class of models and methods for which it is possible 
to write the cost of a segment as follows:
$$ \CostSegmentk = \min_{\mu} \left\{ \SumSegmentkDataPoint \Gammai + g(\mu) \right\},$$
where $\gamma$ is a loss function, depending on the data-point $Y_i$ and 
the (possibly multi-dimensional) parameter $\mu$, and the function $g$ is a regularization penalty. 
\modif{}{For this class of models we show in the following that 
it is theoretically possible to prune the set of segmentations and 
we explain how to implement this pruning if $\mu$ is unidimensional.}

The quadratic loss belongs to this class of model: $\Gammai = (Y_i - \mu)^2$.
This framework also includes maximum likelihood inference of identically distributed and independent data-points. 
Indeed, it suffices to take $\gamma$ equal to minus the log-likelihood: $\Gammai = -\log(p(Y_i, \mu))$.
\modif{}{Let us give some examples. 
\begin{enumerate}
\item We can consider a linear model. To do that we take
$Y_i = (Z_i, x_{i1}, \ldots, x_{ip})$ in $\mathbb{R}^{p+1}$, $\mu = (\beta_0, \beta_1, \ldots, \beta_p, \sigma^2)$ in $\mathbb{R}^{p+1}\times \mathbb{R}^+$ 
and within a segment
$\tau_{k-1}:\tau_{k}$ we assume that
\begin{eqnarray*}
Z_i = \beta_0 + \sum_{j=1}^p \beta_{j}  x_{ij} + \varepsilon_i \quad \text{with} \quad \varepsilon_i \sim \mathcal{N}(0, \sigma^2) \quad \text{i.i.d.}
\end{eqnarray*}
In that case minus the log-likelihood is 
$$
\Gammai = \frac{1}{2} \log(2\pi\sigma^2) + \frac{1}{2\sigma^2} (Z_i -  \beta_0 - \sum_{j=1}^p \beta_{j}  x_{ij})^2.
$$
If the variance is known this simplifies to
$
\Gammai =   (Z_i -  \beta_0 - \sum_{j=1}^p \beta_{j}  x_{ij})^2. 
$

\item We can also consider the segmentation in the mean of a $p$-dimensional sequence. Here we take
$Y_i = (X_{i1}, \ldots X_{ip})$ in $\mathbb{R}^p$, $\mu = (\beta_1, \ldots, \beta_p, \sigma^2)$ in $\mathbb{R}^{p}\times\mathbb{R}^+$
and for all $i$ within a segment $\tau_{k-1}:\tau_{k}$ we assume that
\begin{eqnarray*}
X_{ij} = \beta_j + \varepsilon_{ij} \quad \text{with} \quad \varepsilon_{ij} \sim \mathcal{N}(0, \sigma^2) \quad \text{i.i.d.}
\end{eqnarray*}
In that case minus the log-likelihood is
$$
\Gammai = \frac{p}{2} \log(2\pi\sigma^2) + \frac{1}{2\sigma^2} \sum_{j=1}^p (X_{ij} -  \beta_j )^2.
$$
If the variance is known this simplifies to
$
\Gammai =  \sum_{j=1}^p (X_{ij} -  \beta_j )^2. 
$

\item We could also consider categorical data. Indeed, we can take
$Y_i \in \{1, \ldots, p\}$, $\mu = (\pi_1, \ldots, \pi_p)$ in $[0, 1]^p$ with $\sum_j \pi_j = 1$ and assume that within a segment, 
$Y_i$ are independent and follow a multinomial distribution with
$P(Y_i = j) = \pi_j.$
In that case minus the log-likelihood is $$
\Gammai = -\sum_{j=1}^p \mathbb{I} \{ Y_i = j\} \log(\pi_j),
$$ where $\mathbb{I}$ is the indicator function.
\end{enumerate}
}

Many segmentation models do not include a regularization penalty ($g(\mu) = 0$), however, in some cases
it can be of interest to include a regularization penalty such as the ridge penalty. 
We could take this into account by setting $g(\mu) = \lambda\mu^2$. For simplicity, in the rest of this paper we will only consider the case $g(\mu)=0$, however
extension to $g(\mu) \neq 0$ is straightforward.


\section{Exact segment neighbourhood algorithm and pruning}\label{section:pre_algo}

In this section, we first describe the main update rule of the standard dynamic programming algorithm
- often called the segment neighbourhood algorithm - used to recover the segmentations of minimal cost for $K=1$ to $K_{max}$:

\begin{equation}\label{eq:optimalCost}
\Cost_{\OnetoN+1}^K = \min_{m \in \Mc_{\OnetoN+1}^K} \{ R_m \}, 
\end{equation}

Then we present the functional cost representation of a segmentation and
informally explain how using this representation it is possible to prune 
the search space of the segment neighbourhood algorithm.
In the last subsection (\ref{subsection:IQFP}) we \modif{}{explicitly describe the optimal partitioning problem and 
the advantages and disadvantages of
functional pruning over inequality based pruning (as is done in PELT \citep{killick2012optimal}).}


\subsection{The segment neighbourhood algorithm update rule}
$R_m$ is segment additive, thus the Bellman optimality principle applies and 
if a segmentation is optimal any sub-segmentation of this segmentation is also optimal.
Mathematically, this can be expressed as the following update rule:
\begin{equation}\label{eq:updateSegAdd}
\Cost_{\OnetoT}^K = \min_{\tau < t} \{ \Cost_{\OnetoTau}^{K-1} + \CostSegmentTauT \}, 
\end{equation}
This update rule can be performed with an $\Oc(t)$ \modif{}{time} complexity.
Many algorithms developed for various statistical models implement this particular update rule
\citep{fisher1958grouping,bellman1961approximation,auger1989algorithms,BP03,guedon2008exploring}.
This algorithm was called  the segment neighbourhood algorithm by \citet{auger1989algorithms}.
To recover $\Cost_{\OnetoN +1}^K$ we need to apply update rule (\ref{eq:updateSegAdd}) for every $t$ smaller than $n+1$ and for every $K$ smaller than $\Kmax$.
The overall \modif{}{time} complexity is thus in $\Oc(\Kmax n^2)$. Most of these algorithms have
an $\Oc(n^2)$ space complexity. However some, like \citet{guedon2008exploring}, have an $\Oc(\Kmax n)$ space complexity.

\subsection{Functional cost}
The segment neighbourhood algorithm and in fact most algorithms and heuristics that aim at recovering the 
segmentation of minimal cost 
operate on segmentations through their costs, $R_m$, which are numbers in $\mathbb{R}$. 
Our pruned DPA algorithm is different and uses a slightly more complex representation that we 
call the functional cost of a segmentation. 
We define this function of $\mu$ as:
\begin{equation}\label{eq:functionalCost}
\widetilde{R}_m(\mu) =  \SumSegmentsZeroKminusOne \CostSegmentk + \ \FCostSegmentK(\mu)
\end{equation}
\begin{eqnarray*}
\mbox{with}  \ \FCostSegmentK(\mu) = & \SumSegmentKDataPoint \Gammai & \mbox{if} \ \tau_{K+1} > \tau_K \\
\mbox{and}  \  \FCostSegmentK(\mu)= & 0 & \mbox{if} \ \tau_{K+1} = \tau_K
\end{eqnarray*}
In words, the functional cost, $\widetilde{R}_m(\mu)$, is the cost of segmentation $m$ if the parameter of 
$m$'s last segment is set to $\mu$, rather than the optimal value $\hat{\mu}=\argmin\{ \FCostSegmentK(\mu)\}$.
The minimum value of the functional cost is simply the cost: {\it{i.e.}} $\min_{\mu} \{ \widetilde{R}_m(\mu) \} = R_m$

The functional cost is a more complex representation of a segmentation than the cost, however,
 it leads to some great simplifications. Namely the functional 
cost is point additive, {\it{i.e}} if we consider a segmentation $m$ of $\Mc_{\OnetoT}^K$ with change-points 
$\tau_1, \tau_2, ..., \tau_K$ and the segmentation $m'$ of $\Mc_{\OnetoT+1}^K$ with the same change-points
we have: 
$$\widetilde{R}_{m'}(\mu) = \widetilde{R}_m(\mu) + \gamma(Y_{t}, \mu).$$

In the next sub-section (\ref{subsection:pruning}) we explain informally how this functional formulation 
and the point additiveness makes it theoretically possible to prune the set of segmentations.

\subsection{Functional cost and pruning}\label{subsection:pruning}
\modif{}{T}he pDPA search\modif{}{es} for:
\begin{equation}\label{eq:optimalFunCost}
\FCost_{\OnetoN+1}^K(\mu) = \min_{m \in \Mc_{\OnetoN+1}^K} \{ \widetilde{R}_m(\mu) \}, 
\end{equation}
for $K=1$ to $K_{max}$. \modif{}{This is the functional formulation of equation~(\ref{eq:optimalCost}).}
Like the cost \modif{}{(see update rule~(\ref{eq:updateSegAdd}))}, the functional cost is segment additive and a similar update rule applies:
\begin{equation}\label{eq:FirstUpdateOptimalFunCost}
\FCost_{\OnetoT}^K(\mu) = \min_{\tau < t} \{ \Cost_{\OnetoTau}^{K-1} + \FCostSegmentTauT(\mu)  \}.
\end{equation}

Thanks to the point-additiveness of $\widetilde{R}_m(\mu)$, it is possible to further simplify this update rule.
If we consider a given value of $\mu$, two last change-points $\tau$ and  $\tau'$ 
and a time $t'>t$ we get the following implication:
\begin{equation}\label{eq:simplepruning}
\left\{ \Cost_{\OnetoTau}^{K-1} + \FCostSegmentTauT(\mu) \leq \Cost_{\OnetoTau'}^{K-1} + \FCostSegmentTauprimeT(\mu) \right\}
 \ \implies \ 
\left\{ \Cost_{\OnetoTau}^{K-1} + \FCostSegmentTauTprime(\mu) \leq \Cost_{\OnetoTau'}^{K-1} + \FCostSegmentTauprimeTprime(\mu)\right\}.
\end{equation}
In other words, if for a given $t$ and $\mu$ the functional cost of 
last change-point $\tau$ is less than the functional cost of $\tau'$, then this will still be the case 
for any time point in the future ({\it{i.e.}} for any $t'>t$).
Hence for the parameter value $\mu$ we can discard $\tau'$ as it will never be minimal. 

If we consider only one value for $\mu$, this pruning rule can be formalised as the following update rule:
\begin{equation}\label{eq:SecondUpdateOptimalFunCost}
\FCost_{\OnetoT+1}^K(\mu) = \min \left\{ \FCost_{\OnetoT}^K(\mu) + \gamma(Y_{t}, \mu) \ , 
\  \Cost_{\OnetoT}^{K-1} + \gamma(Y_{t}, \mu) \right\},
\end{equation}
and this update rule can be performed with an $\Oc(1)$ \modif{}{time} complexity\modif{}{. We only need to
compare the values of }
$\FCost_{\OnetoT}^K(\mu) + \gamma(Y_{t}, \mu)$ and $\Cost_{\OnetoT}^{K-1} + \gamma(Y_{t}, \mu)$ \modif{)}{}.


\subsection{Implementing update rule (\ref{eq:SecondUpdateOptimalFunCost}) for all possible $\mu$}\label{subsection:allmu}
\begin{figure}[H]
\begin{center}
\parbox{0.3cm}{\begin{turn}{90} $Y_i$\end{turn}
}\parbox{5.5cm}{
\includegraphics[scale=0.3]{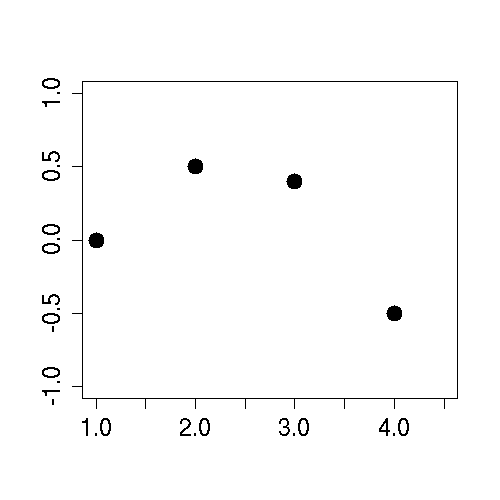} 
\vspace{-1cm}
\begin{center}$i$\end{center}
}
\end{center}
\caption{Four-point signal. $y_i$ as a function of $i$. $y_1 = 0, \ y_2 = 0.5,\ y_3= 0.4,\ y_4= -0.5$}\label{fig:data}
\end{figure}

Update rule (\ref{eq:SecondUpdateOptimalFunCost}) cannot be implemented in practice, 
because the rule would have to be applied to all possible values of $\mu$ 
and in most cases the set of all $\mu$ is uncountably infinite. 
A key idea to cope with this problem is to consider the set of $\mu$ for which a particular last change-point $\tau$
is better than \modif{}{any other change-point $\tau'$ in the sense that $ \Cost_{\OnetoTau}^{K-1} +\FCostSegmentTauT(\mu)$ is smaller than  $ \Cost_{\OnetoTau'}^{K-1} + \FCostSegmentTauprimeT(\mu)$}. We will call this set $ \Sc_{\OnetoT,\tau}^K$.
We will properly define and study the properties of $\Sc_{\OnetoT, \tau}^K$ in section \ref{section:algo}. 
As yet let us have a look at these sets on a small example. 
We will consider the segmentations with one change-point ($K=1$) of a four-point signal with the quadratic loss: $\gamma(y_i, \mu)= (y_i - \mu)^2$.
Our observations are $y_1 = 0, \ y_2 = 0.5,\ y_3= 0.4,\ y_4= -0.5$ (see Figure \ref{fig:data}). 

\begin{figure}[h]

\parbox{0.3cm}{\begin{turn}{90} Functional Cost\end{turn}
}\parbox{5cm}{
\includegraphics[trim= 10mm 0mm 0mm 0mm, clip=true, scale=0.3]{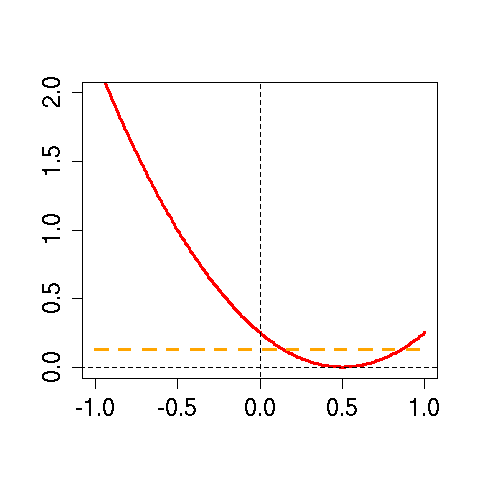}
\vspace{-1cm}
\begin{center}$\mu$\end{center}
}\parbox{5cm}{

\renewcommand{\arraystretch}{1.3}
\begin{tabular}{|c|c|c|}
\hline
$\tau$ & Functional cost &  $\Sc_{1:3, \tau}^1$  \\
\hline
$2$ & \textcolor{red}{$0.25 - \mu + \mu^2$}  & $ [\ 0.146 \ , \  0.854 \ ]$  \\
\hline 
$3$ & $ \textcolor{orange}{0.125} $  &  $[ \ -\infty , 0.146 \ ] \cup [ \ 0.854 , +\infty \ ]$  \\
\hline 
\end{tabular}
}

\caption{Functional cost of $Y_{1:3}$ for $K=1$  using the quadratic loss. 
(Left) Functional cost as a function of $\mu$ of segmentations having a change-point at $\tau=2$ (solid red) 
and $\tau=3$ (orange dashed). (Right) Analytical expression of the functional costs 
for $\tau=2$ and $\tau=3$ and the set of $\mu$, 
for which they are optimal: $\Sc_{1:3, \tau}^1$.}\label{fig:step2}
\end{figure}

\paragraph{Functional cost of $Y_{1:3}$} 
Let us first consider the first two data-points: $Y_{1:3}$.
The first segmentation we need to consider is the one with a change-point at $\tau=2$.
The functional cost of this segmentation is simply the cost of its first segment,
$r_{1:2}$, which is $0$, plus the functional cost of the last segment, 
which is the polynomial function $\mu \rightarrow (y_2 - \mu)^2$. 
We should also consider a segmentation with a change-point at $\tau=3$.
Indeed, for $t=2$ its functional cost is well define\modif{}{d}:
it is the cost of its first segment, $r_{1:3}$, which is $0.125$, plus the functional cost
of its last segment, $r_{3:3}$. $r_{3:3}$ is empty and so its functional cost is simply the zero function: $\mu \rightarrow 0$.
These two functional costs are represented in figure \ref{fig:step2}-left 
as a solid red line for $\tau=2$ and a dashed orange line for $\tau=3$. 
Simple \modif{}{} calculations show  that the set of $\mu$ for which $\tau=2$ is \modif{}{best}, 
$\Sc_{1:3,2}^1$, is an interval cent\modif{}{e}red on $0.5$: $[\ 0.146 \ , \  0.854 \ ]$. $\Sc_{1:3,3}^1$ is a union of two intervals: 
$[ \ -\infty , 0.146 \ ] \cup [ \ 0.854 , +\infty \ ]$. 
These intervals are also given in figure \ref{fig:step2}-right.

\begin{figure}[H]
\begin{center}
\parbox{0.3cm}{\begin{turn}{90} Functional Cost\end{turn}
}\parbox{5cm}{
\includegraphics[trim= 10mm 0mm 0mm 0mm, clip=true, scale=0.3]{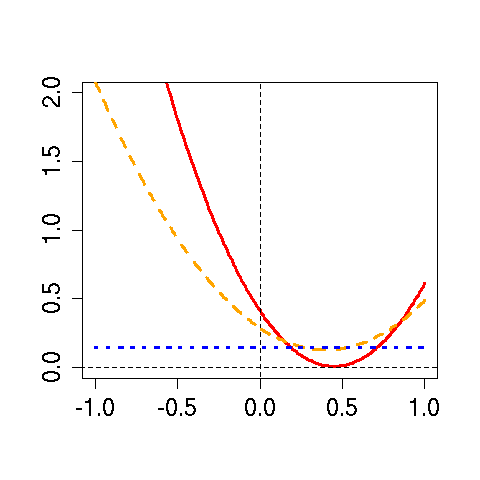}
\vspace{-1cm}
\begin{center}$\mu$\end{center}
}\parbox{6.5cm}{

\renewcommand{\arraystretch}{1.3}
\begin{tabular}{|c|c|c|}
\hline
$\tau$ & Functional cost &  $\Sc_{1:4, \tau}^1$  \\
\hline
$2$ & \textcolor{red}{$0.41 - 1.8 \mu + 2 \mu^2$}  & $ [\ 0.190\ , \  0.709 \ ]$  \\
\hline 
$3$ & $ \textcolor{orange}{0.285 - 0.8 \mu + \mu^2} $  &  $\emptyset$  \\
\hline 
$4$ & $ \textcolor{blue}{0.14} $  &  $[ \ -\infty , 0.190 \ ] \cup [ \ 0.709 , +\infty \ ]$  \\
\hline
\end{tabular}
}
\caption{Functional cost of $Y_{1:4}$ for $K=1$ \modif{}{} using the quadratic loss. 
(Left) Functional cost of a segmentations having a change-point at $\tau=2$ (solid red) 
$\tau=3$ (orange dashed) and $\tau=4$ (blue \modif{}{dotted}). (Right) Analytical expression of the functional costs 
for $\tau=2$, $3$ and $4$ and the set of $\mu$, 
for which they are optimal: $\Sc_{1:4, \tau}^1$.}\label{fig:step3}
\end{center}
\end{figure}

\paragraph{Functional cost of $Y_{1:4}$} 
Now if we consider the first three data-points: $Y_{1:4}$.
We have to update the functional cost of the segmentations having a change-point at $\tau=2$ and $3$.
We do this by adding the function $\mu \rightarrow (y_3 - \mu)^2$ to both of them (point additiveness). We also need to consider another possible 
change-point: $\tau=4$. Its functional cost is simply the cost of segment $r_{1:4}$, which is $0.14$,
plus the function cost of segment $r_{4:4}$ which is the zero function: $\mu \rightarrow 0$.
These three functional costs are represented in figure \ref{fig:step3}-left 
as a solid red line for $\tau=2$, a dashed orange line for $\tau=3$ and a blue \modif{}{dotted} line for $\tau=4$. 
Simple calculations show  that the set of $\mu$ for which $\tau=2$ is best, 
$\Sc_{1:4,2}^1$, is an interval cent\modif{}{e}red on $0.45$: $[\ 0.190\ , \  709 \ ]$. $\Sc_{1:4,4}^1$ is a union of two intervals: 
$[ \ -\infty , 0.190 \ ] \cup [ \ 0.709 , +\infty \ ]$. 
$\Sc_{1:4,3}^1$ is strikingly empty. Based on equation (\ref{eq:simplepruning}), this means that irrespective of the last data-point it is sure that the segmentation
with a last change-point at $\tau=3$ will always have a greater cost that those with a change-point at $2$ or $4$.
Thus for all possible value\modif{}{s} of $\mu$ we can discard $\tau=3$.
Note however that neither $\tau=2$ nor $\tau=4$ are better than $\tau=3$ for all $\mu$.

This simple example illustrates that by keeping track of the $\Sc_{\OnetoT, \tau}^K$
we might be able to prune some candidate change-points very early in the process. 
We will make this idea explicit in the next sections (section \ref{section:algo} and 
\ref{section:complexity}).

\subsection{Segment Neighbourhood, Optimal partitioning and PELT}\label{subsection:IQFP}

\modif{}{
Since 2010 another pruned algorithm, PELT, has been proposed for change-point detection problems by \citet{killick2012optimal}.
pDPA and PELT are different in nature. 
First, they do not solve the same problem.
PELT is an extension of the optimal partitioning algorithm \citep{jackson2005algorithm} 
rather than the segment neighbourhood algorithm.
More precisely, if we call $\Mc_{\OnetoN+1}$ the set of all segmentations, the optimal partitioning problem is
to find:}
\begin{equation}\label{eq:OPCost}
\OPCost_{\OnetoN+1} = 
\min_{m \in \Mc_{\OnetoN+1}} \{ R_m + \lambda |m| \}, 
\end{equation}
\modif{}{where $R_m$ is defined in equation (\ref{eq:SegAddCost}), 
$\lambda$ is a user defined scalar and $|m|$ is the number of change-points of segmentation $m$. 
Intuitively, the number of change-points of the recovered solution decreases with $\lambda$.
If the penalty $\lambda$ is known in advance solving the optimal partitioning problem is faster 
than solving the segment neighbourhood problem.
The advantage of solving the segment neighbourhood problem is that this gives optimal segmentations 
for a range of numbers of change-points. 
}

\modif{}{Second PELT and the pDPA do not prune segmentations in the same way. PELT's prune segmentations
based on their cost  and the pDPA based on their functional cost. In a recent pre-print,
\citet{maidstone2014optimal}
called these two ways of pruning  respectively 
inequality based pruning (IP) and functional based pruning (FP) and studied their relationship.
The benefit of IP over FP is that it is more widely applicable than FP, i.e. the assumptions on 
the segment cost to apply FP imply the assumptions to apply IP. Furthermore the implementation of 
IP is usually straightforward which is not the case of FP.
The advantage of FP over IP is that it can prune efficiently the set of change-points even if there are
only a few change-points in the sequence. We prove this for a special loss function
and under some restrictive conditions in sub-section \ref{subsection:Averagecomplexity}. 
We empirically confirm this result for the quadratic loss in 
sub-section \ref{subsection:empiricalcomplexity}.
Furthemore,  
\citet{maidstone2014optimal}
proved in Theorem 6.1 for the optimal partitioning and segment neighbourhood problem that any candidate change-point pruned
by IP will also be pruned by FP at worst at the same time. In other words FP cannot prune less than IP.
}

\section{Functional pruning of the set of segmentations}\label{section:algo}
In this section we first describe the key quantities and properties used by the pruned DPA algorithm. 
Then we describe the algorithm.


\subsection{Key quantities and their properties}\label{subsection:property}
The four main quantities of interest in the pDPA are given below.
\begin{enumerate}
\item the optimal functional cost with $K$ change-points:
$$\FCost_{\OnetoN+1}^K(\mu) = \min_{m \in \Mc_{\OnetoN+1}^K} \{ \widetilde{R}_m(\mu) \},$$
where the functional cost of a segmentation $\widetilde{R}_m(\mu)$ is defined in equation
(\ref{eq:functionalCost}).

\item the optimal functional cost with $K$ change-points if the last change-point is $\tau \leq t$: 
\begin{equation} \FCc_{\OnetoT, \tau}^K(\mu) = \Cost_{\OnetoTau}^{K-1} + \FCostSegmentTauT(\mu),\label{eq:functionalCostChange}
\end{equation}
where $\Cost_{\OnetoTau}^{K-1}$ is the optimal cost with $K-1$ change-points up to $\tau$ defined in 
equation (\ref{eq:optimalCost}) and $\FCostSegmentTauT(\mu)$ 
is defined in equation (\ref{eq:functionalCost}).

\item the set of $\mu$ for which the last change-point $\tau$ is optimal:
$$ \Sc_{\OnetoT, \tau}^K = \{ \mu \ | \FCc_{\OnetoT, \tau}^K(\mu) = \FCost_{\OnetoT}^K(\mu) \}.$$

\item the set of last change-points which are optimal for at least one $\mu$
\footnote{\label{1}Depending on the properties of 
$\gamma$ it might be possible to consider only the set of last change-points $\tau$ for which
 $\Sc_{\OnetoT, \tau}^K$ is not restricted to a finite set of $\mu$. This is the case for the quadratic loss which is continuous.\label{footnote:1}} 
$$ \Tauc_{\OnetoT}^{K} = \left\{ \tau < t \ | \ \Sc_{\OnetoT, \tau}^K \neq \emptyset \right\}$$

\end{enumerate}

Below are the properties of these quantities.
The proofs of these properties are given afterwards.


\begin{prop} $ $ \label{prop:funSegAdd}
\modif{}{}
\modif{}{The optimal functional cost may be computed using only the (pruned) set of change-points 
which are optimal for at least one $\mu$.}
\begin{eqnarray*} 
\FCost_{\OnetoT}^K(\mu) = & \underset{\tau < t}{\min} \{ \FCc_{\OnetoT, \tau}^K(\mu) \} \\
 = & \underset{\tau \in \Tauc_{\OnetoT}^{K}}{\min} \{ \FCc_{\OnetoT, \tau}^K(\mu) \}.
\end{eqnarray*}
\end{prop}

\begin{prop} $ $ \label{prop:funCost}
\modif{}{}
\modif{}{The functional cost with a last change-point $\tau$ is easy to update  using point additivity.}
\begin{eqnarray*} 
\FCc_{\OnetoT+1, \tau}^K(\mu) =  \FCc_{\OnetoT, \tau}^K(\mu) + \Gammat.
\end{eqnarray*}
\end{prop}

\begin{prop} $ $ \label{prop:set}
\modif{}{The set of $\mu$ for which a change-point $\tau$ is optimal decreases with $t$ (w.r.t. set inclusion) and can be pruned.}
\begin{eqnarray}\label{eq:set1}
\forall \ \tau < t, \quad \Sc_{\OnetoT+1, \tau}^K & = & \Sc_{\OnetoT, \tau}^K \quad \bigcap \quad \{ \mu \ | \ \FCc_{\OnetoT, \tau}^K(\mu) \leq \Cost_{\OnetoT}^{K-1} \} 
\end{eqnarray}

\begin{eqnarray}\label{eq:set2}
\Sc_{\OnetoT+1, t}^K & = & \bigcap_{\tau \in \Tauc_{\OnetoT}^{K}} \ \{ \mu \ | \ \Cost_{\OnetoT}^{K-1} \leq  \FCc_{\OnetoT, \tau}^K(\mu) \} 
\end{eqnarray}

\begin{equation}\label{eq:set3}
 \Sc_{\OnetoT, \tau}^K = \emptyset \quad \implies \quad \forall \ t'\geq t, \ \Sc_{\OnetoT', \tau}^K = \emptyset.
\end{equation}
\end{prop}

\begin{prop} $ $ \label{Prop:change}
\modif{}{The set of candidate change-points can be pruned and computed recursively.}
\begin{equation*}
\Tauc_{\OnetoT+1}^{K} =  \quad \{ \tau \in (\Tauc_{\OnetoT}^{K}\cup \{t\}) \ | \
\Sc_{\OnetoT+1, \tau}^K \neq \emptyset \}.
\end{equation*}
\end{prop}

Property \ref{prop:funSegAdd} is very similar to the update rule of
the segmentation neighbourhood algorithm (equation (\ref{eq:updateSegAdd})).
The main difference is that the set of last change-points to consider is not necessarily all 
those before $t$: $\{\tau | \tau < t\}$ but rather those that still have a chance to be optimal
at this stage: $\Tauc_{\OnetoT}^{K}$. At worst this $\Tauc_{\OnetoT}^{K}$ is equal to $\{\tau | \tau < t\}$ but based on property \ref{Prop:change} \modif{}{this set could be smaller}. Indeed any change-point that has been discarded from $\Tauc_{\OnetoT}^{K}$ will never be included again.

The four previous properties are in fact simple consequences of the point additiveness and segment additiveness of the functional cost. 
However given their combinatorial nature they might look a bit tedious.
To clarify those properties we provide detailed proofs in the following paragraphs.

\paragraph{Proof of property \ref{prop:funSegAdd}}
As the cost in the segment neighbourhood algorithm, 
the functional cost of a segmentation is segment additive, 
thus the Bellman optimality principle holds and we recover 
the first part of property \ref{prop:funSegAdd}. 
The second part follows by definition of $\Tauc_{\OnetoT}^{K}$. 
Indeed, if $\tau$ is the optimal last change-point then there must exist a $\mu$ for which its functional cost is equal to 
$\FCost_{\OnetoT}^{K}(\mu)$
 and thus $\tau$ is in $\Tauc_{\OnetoT}^{K}$ $\ \blacksquare$

\paragraph{Proof of property \ref{prop:funCost}}
By definition of $\FCc_{\OnetoT, \tau}^K(\mu)$ (see equation (\ref{eq:functionalCostChange})) we have:
\begin{eqnarray*}
\FCc_{\OnetoT+1, \tau}^K(\mu)  = & \Cost_{\OnetoTau}^{K-1} + \ \SumSegmentTautoTplusOne \Gammai & \\
 = &  \Cost_{\OnetoTau}^{K-1} + \ \SumSegmentTautoT \Gammai & + \ \ \Gammat \\
 = & \FCc_{\OnetoT, \tau}^K(\mu) + \ \Gammat  & \ \ \blacksquare
\end{eqnarray*}

\paragraph{Proof of property \ref{prop:set}} 
Using property \ref{prop:funSegAdd} and by definition of $\Sc_{\OnetoT+1, \tau'}^K$ we get that for $\tau' < t+1$:

\begin{eqnarray*}
\Sc_{\OnetoT+1, \tau'}^K = & \left\{ \mu \ | \ \FCc_{\OnetoT+1, \tau'}^K(\mu) \leq \underset{\tau < t+1}{\min} \{ \FCc_{\OnetoT+1, \tau}^K(\mu) \} \right\} \\
                        = & \underset{\tau < t+1}{\bigcap} \left\{ \mu \ | \ \FCc_{\OnetoT+1, \tau'}^K(\mu) \leq  \FCc_{\OnetoT+1, \tau}^K(\mu) \right\}. \\
\end{eqnarray*}

Now, using  property \ref{prop:funCost} we also have for $\tau < t+1$:
\begin{equation*}
\{ \mu \ | \ \FCc_{\OnetoT+1, \tau'}^K(\mu) \leq  \FCc_{\OnetoT+1, \tau}^K(\mu)\} = 
\{ \mu \ | \ \FCc_{\OnetoT  , \tau'}^K(\mu) \leq  \FCc_{\OnetoT  , \tau}^K(\mu)\}.
\end{equation*}
Finally, by defintion we also have $\FCc_{\OnetoT, t}^K(\mu) = \Cost_{\OnetoT}^{K-1}.$

Combining these three facts for $\tau' < t$ we get:
\begin{eqnarray*}
\Sc_{\OnetoT+1, \tau'}^K = & \underset{\tau < t+1}{\bigcap} \ \left\{ \mu \ | \FCc_{\OnetoT, \tau'}^K(\mu) \leq  \FCc_{\OnetoT, \tau}^K(\mu) \right\} \\
                         = & \Sc_{\OnetoT, \tau'}^K \ \bigcap \ \left\{ \mu \ | \ \FCc_{\OnetoT, \tau'}^K(\mu) \leq  \FCc_{\OnetoT, t}^K(\mu) \right\} \\
                         = & \Sc_{\OnetoT, \tau'}^K \ \bigcap \ \left\{ \mu \ | \ \FCc_{\OnetoT, \tau'}^K(\mu) \leq \Cost_{\OnetoT}^{K-1} \right\}
\end{eqnarray*}
and we recover equation (\ref{eq:set1}) of property \ref{prop:set}.

We recover equation (\ref{eq:set2}) of property \ref{prop:set} by taking $\tau' = t$:
\begin{eqnarray*}
\Sc_{\OnetoT+1, t}^K & = & \underset{\tau < t}{\bigcap} \left\{ \mu \ | \ \Cost_{\OnetoT}^{K-1} \leq  \FCc_{\OnetoT, \tau}^K(\mu) \right\} \\
& = & \left\{ \mu \ | \ \Cost_{\OnetoT}^{K-1} \leq \underset{\tau < t}{\min} \{ \FCc_{\OnetoT, \tau}^K(\mu) \} \right\} \\
& = & \left\{ \mu \ | \ \Cost_{\OnetoT}^{K-1} \leq \underset{\tau \in  \Tauc_{\OnetoT}^{K}}{\min} \{ \FCc_{\OnetoT, \tau}^K(\mu) \} \right\} \\
& = & \underset{\tau \in \Tauc_{\OnetoT}^{K}}{\bigcap} \left\{ \mu \ | \ \Cost_{\OnetoT}^{K-1} \leq  \FCc_{\OnetoT, \tau}^K(\mu) \right\}.
\end{eqnarray*}

Finally, if $\Sc_{\OnetoT, \tau}^K=  \emptyset$ using equation (\ref{eq:set1}) we have that
$$\Sc_{\OnetoT+1, \tau}^K = \Sc_{\OnetoT, \tau}^K \ \bigcap \ \{ \mu \ | \ \FCc_{\OnetoT, \tau}^K(\mu) \leq \Cost_{\OnetoT}^{K-1} \} = \emptyset$$ 
and by induction we recover equation (\ref{eq:set3}) of property \ref{prop:set}$\ \blacksquare$

\paragraph{Proof of property \ref{Prop:change}}
By definition of $\Tauc_{\OnetoT+1}^{K}$ we have 
$$ \Tauc_{\OnetoT+1}^{K} \quad \supseteq  \quad \{ \tau \in (\Tauc_{\OnetoT}^{K}\cup \{t\}) \ | \
\Sc_{\OnetoT+1, \tau}^K \neq \emptyset \}.$$

Now suppose $\tau$ is in $\Tauc_{\OnetoT+1}^{K}$. If $\tau < t$, then using equation 
(\ref{eq:set3}) of property \ref{prop:set} we see that $\tau$ must also be in $\Tauc_{\OnetoT}^{K}$.
If $\tau=t$, then by definition of $\Tauc_{\OnetoT+1}^{K}$ we have that $\Sc_{\OnetoT+1, t}^K$ is not empty. Thus we recover:
$$ \Tauc_{\OnetoT+1}^{K} \quad \subseteq \quad  \{ \tau \in (\Tauc_{\OnetoT}^{K}\cup \{t\}) \ | \
\Sc_{\OnetoT+1, \tau}^K \neq \emptyset \} \ \blacksquare$$

\subsection{The pDPA algorithm}
The pruned dynamic programming algorithm use\modif{}{s} the four properties described in 
subsection \ref{subsection:property} to update the optimal functional cost, 
the set of optimal last change-points and the set of $\mu$ for which these last change-points are optimal. 
For every $t \leq n$ and $K \leq \Kmax$ the algorithm:
\begin{enumerate}
\item updates for each $\tau$ in $\Tauc_{\OnetoT}^{K}$ the set of $\mu$ for which they are optimal (equation (\ref{eq:set1}));
\item initialises the set of value\modif{}{s} for which  a change-point at $t$ is optimal (equation (\ref{eq:set2}));
\item updates the set of possible last change-points (property \ref{Prop:change});
\item updates the functional cost of possible last change-points (property \ref{prop:funCost});
\item computes the minimal cost with $K$ change-points at $t$ (property \ref{prop:funSegAdd}).
\end{enumerate}
More formally the algorithm can be described as:

\begin{algorithm}[H]
 \KwData{A sequence $Y_{\OnetoN+1}$}
 \KwResult{Two matrices $D_{K,t}$ and  $I_{K,t}$ of size $(n+1) \times \Kmax$ containing the optimal cost and optimal last change-points }
 \For{$K$ $=$ $1$ to  $\Kmax$}
 {
	\vspace{0.1cm}
	$\Tauc_{1:K+1}^{K} \leftarrow \{K\}$

	$\FCc_{1:K+1, K}^K(\mu) \leftarrow \Cost_{1:K}^{K-1} + \gamma(Y_K,\mu)$

	\vspace{0.1cm}

 	\For{$t$ $=$ $K+1$ to  $n$}
	{
	\vspace{0.2cm}
    		\For{$\tau$ $\in$ $\Tauc_{\OnetoT}^{K}$}
		{
    			$\Sc_{\OnetoT+1, \tau}^K  \leftarrow  \Sc_{\OnetoT, \tau}^K \quad \bigcap \quad 				\{ \mu \ | \ \FCc_{\OnetoT, \tau}^K(\mu) \leq \Cost_{\OnetoT}^{K-1} \}$
 		}

		\vspace{0.1cm}
		$\Sc_{\OnetoT+1, t}^K  \leftarrow  \quad \bigcap_{\tau < t} \ \{ \mu \ | \ 
		\Cost_{\OnetoT}^{K-1} \leq \FCc_{\OnetoT, \tau}^K(\mu) \}$

		\vspace{0.1cm}

		$\Tauc_{\OnetoT+1}^{K} \leftarrow \quad \{ \tau \in (\Tauc_{\OnetoT}^{K}\cup \{t\}) \ | \
		\Sc_{\OnetoT+1, \tau}^K \neq \emptyset \}$

		\vspace{0.1cm}
		\For{$\tau$ $\in$ $\Tauc_{\OnetoT+1}^{K}$}
		{
    			$\FCc_{\OnetoT+1, \tau}^K(\mu) \leftarrow  \quad \FCc_{\OnetoT, \tau}^K(\mu) + \Gammat$	
 		}
		\vspace{0.1cm}
		$D_{K,t+1} \leftarrow  \quad \underset{\tau \in \Tauc_{\OnetoT+1}^{K}}{\min} 
			\left\{ \min_{\mu}\{ \FCc_{\OnetoT+1, \tau}^K(\mu) \} \right\}$

		$I_{K,t+1} \leftarrow \quad \underset{\tau \in \Tauc_{\OnetoT+1}^{K}}{\argmin} 
			\left\{ \min_{\mu}\{ \FCc_{\OnetoT+1, \tau}^K(\mu) \} \right\}$
 	}
	\vspace{0.2cm}
 }
 \caption{Pruned DPA algorithm}
\end{algorithm}

Importantly $\FCost_{\OnetoT}^{K}(\mu)$, $\Sc_{\OnetoT, t}^K$ and $\Tauc_{\OnetoT}^{K}$
are used only at step $K, t$ and $K, t+1$ \modif{}{. So they} need not be stored, they can be discarded or overwritten immediately.

\paragraph{Implementing the pruned DPA}
The pruned DPA critically relies on the possibility of easily updating the functional cost 
($\mu \rightarrow \FCc_{\OnetoT, \tau}^K(\mu)$) and the set of $\mu$ for which one particular last change-point is optimal 
($\Sc_{\OnetoT, \tau}^K$). If this is not the case then the algorithm and the fact
that the set of last change-points can be functionally pruned is purely theoretical.

Representing and updating the functional cost is easy typically if there exists a simple analytical decomposition of the functional cost. For example, if we consider the quadratic loss, the functional cost is a second degree polynomial function and it can be represented by the three coefficients of this polynomial function. 
Other loss functions can be represented in such a way, and in fact the pruned DPA has been implemented for other losses since its first pre-print (namely the Poisson and 
negative binomial log-likelihood losses, \citet{cleynen2014segmentor3isback})

Representing the sets $\Sc_{\OnetoT, \tau}^K$ is \modif{}{a priori more difficult}. Their representation depends on the dimensionality of $\mu$ and the characteristic of the loss function $\gamma$.
\modif{

}{
In the next section we study theoretically and empirically the expected and worst case complexity of functional pruning in the simple case where $\mu$
is a unidimensional parameter in $\mathbb{R}$ and all functions $\mu \rightarrow \sum \Gammai$ are unimodals.
Here by unimodal we mean functions such that for any $c$ in $\mathbb{R}$ the set $\{ x | f(x) \leq c\}$ is an interval. 
In that case it is simpler to keep track of the sets $\{ \mu \ | \ \Cost_{\OnetoT}^{K-1} \leq  \FCc_{\OnetoT, \tau}^K(\mu) \}$ and $\Sc_{\OnetoT, \tau}^K$ as they are respectively intervals and union of intervals. 
}
\modif{}{Functional pruning is theoretically possible even if those conditions are not valid and in particular if $\mu$ is multidimensional. 
However the implementation and the theoretical study of functional pruning in a multidimensional case is not straightforward and is outside the scope of this paper.}



\section{Complexity of the algorithm for a unidimensional $\mu$}\label{section:complexity}

In this section we study the complexity of the pruned DPA 
for $\mu$ in $\mathbb{R}$. We also assume that updating the 
functional cost is in $\Oc(1)$, which is
the case if there is a simple analytical decomposition of the functional cost.
Finally we will assume that any function $\mu \rightarrow \sum_i \gamma(Y_i, \mu)$
is a unimodal function of $\mu$. This last assumption is true if $\gamma$ is convex, which is  often the case if we take $\gamma$ 
to be minus the log-likelihood. Under those conditions all $\Sc_{\OnetoT, \tau}^K$ are unions of 
intervals\footnote{Indeed, if all $\mu \rightarrow \sum_i \gamma(Y_i, \mu)$ are unimodal functions 
then all $\{ \mu \ | \ \FCc_{\OnetoT, \tau}^K(\mu) \leq \Cost_{\OnetoT}^{K-1} \}$ are intervals 
as $\Cost_{\OnetoT}^{K-1}$ is a constant in $\mathbb{R}$ and all $\FCc_{\OnetoT, \tau}^K(\mu)$ are unimodal. 
Then using property \ref{prop:set} we get by induction that all
$\Sc_{\OnetoT, \tau}^K$ are unions of intervals.}.

A key question then is how many intervals do we precisely need. 
In this section, we propose a bound of this number and using it we bound the worst case
complexity of the pDPA and show that it is at worst as efficient as the segment neighbourhood algorithm (see subsection \ref{subsection:worst}).
Then we theoretically study the average complexity of the pDPA algorithm for a particular loss
function for a random sequence without change-points (see subsection \ref{subsection:Averagecomplexity}).
Finally, we empirically study the complexity of the algorithm for the quadratic loss
(see subsection \ref{subsection:empiricalcomplexity}).

\subsection{Worst case complexity of the pDPA}\label{subsection:worst}

In this section we prove that the pDPA is at worst as efficient as the segment neighbourhood algorithm. More precisely we have the following proposition.

\begin{propo}\label{theo:wrst}
If all $\sum_{j}^{} \gamma(Y_{j}, \mu)$ are unimodal in $\mu$ and if both minimising $\FCc_{\OnetoT, \tau}^K(\mu)$ 
and finding the roots of $\FCc_{\OnetoT, \tau}^K(\mu) \ = \ \Cost_{\OnetoT}^{K-1}$ are in $\mathcal{O}(1)$, 
the pDPA is at worst in $\mathcal{O}(\Kmax n^2)$ time and in $\Oc(\Kmax n)$ space. 
\end{propo}

\paragraph{Proof.}
The key quantity to control is the number of intervals 
needed to represent $\Sc_{\OnetoT, \tau}^K$.
For a given $K$ and at step $t$ the number of candidate last change-points is obviously bound\modif{}{ed} by $t$. 
If all $\sum_{j=\tau+1}^{t+1} \gamma(Y_{j}, \mu)$ are unimodal, using theorem \ref{Lem:Order1} (proved in appendix \ref{Section:LemmaProof}) 
we get that the total number of intervals is bounded by $2t-1$. 
Thus at each step there is at most $t$ last change-points and $2t-1$ intervals to update.
By summing all these bounds from $1$ to $n$ and for every possible $K$ we retrieve an $\mathcal{O}(\Kmax n^2)$ worst case time complexity. 

\noindent As for the worst case space complexity, we need to store two $(n+1)\times \Kmax$ matrices ($D_{K,t}$ and $I_{K,t}$) and at each step 
there is at most $t$ candidates and $2t-1$ intervals. This gives an $\Oc(\Kmax n)$ space complexity~$\ \blacksquare$

\paragraph{}
The previous theorem provides a worst case bound on the complexity.
One can wonder in which cases this quadratic complexity in $n$ is reached.
Let us consider the sequence such that $\forall i, \ Y_i = i$, segmentations with one change-point and the quadratic loss.
The best segmentation of $Y_{\OnetoT}$ \modif{}{h}as a change-point at $t/2$. So at step $t$ of the pDPA
the change-point $\tau = t/2$ has not been pruned and in fact all change-points 
from $t/2$ to $t$ haven't been pruned. Hence, at each step we have at least $t/2$ candidates to 
update. Hence by summing from $t=1$ to $n$ we recover a quadratic complexity: $\Oc(n^2)$.
 
However, as illustrated with an example in subsection \ref{subsection:allmu} and demonstrated
for a specific loss function in the following subsection (\ref{subsection:Averagecomplexity}) 
one can hope that in many cases the pruning will be more efficient than that.


\subsection{Average pruning of the pDPA with a special loss function and $\Kmax=1$}\label{subsection:Averagecomplexity}

In this subsection, we prove for a particular loss function and $\Kmax=1$ that on average we get 
an efficient pruning with the functional cost representation even when we consider a random 
sequence without true change-points. As explained in the introduction this result is not intuitive.
In subsection \ref{subsection:empiricalcomplexity} we empirically show that this is also the case for the quadratic loss function.

\paragraph{}
Here, we will use the negative log-likelihood loss of a continuous uniform distribution defined on $[0, \mu]$.
This loss function is:
\begin{eqnarray*}
\begin{cases}
\gamma(Y_i, \mu) =  \log(\mu) &\mbox{if} \ 0 \leq Y_i \leq \mu \\
\gamma(Y_i, \mu) =  \infty & \mbox{otherwise}
\end{cases}
\end{eqnarray*}
For this particular loss function \footnote{note that any $\mu \rightarrow \sum_i \gamma(Y_i, \mu)$ is unimodal.} 
and for $\Kmax=1$, it is possible to bound the average number of candidate last change-points, $E(|\Tauc_{\OnetoT}^{K}|)$. 

\begin{prop}\label{prop:avt}
For the negative log-likelihood loss, $\Kmax=1$, and 
for, $Y_{\OnetoN+1}$, $n$ independent and identically distributed random variables 
of density $f$ and continuous distribution $F$, $E(|\Tauc_{\OnetoN}^{1}|) = \Oc(\log(n))$ and the average time complexity 
of the pDPA is in $\Oc(n\log(n))$. 
\end{prop}

\paragraph{Proof}
The proof of $E(|\Tauc_{\OnetoT}^{1}|) = \Oc(\log(t))$ is given in appendix \ref{app:averagecomp}.
We obtain this result by studying the set $\Sc_{\OnetoN, \tau}^1$. 
More precisely we characterize some simple events for which $\Sc_{\OnetoN, \tau}^1$ is empty and 
compute the probability of these events. Then by taking the expectation and summing over all possible $\tau$ we get the desired result.

For the complexity using theorem \ref{Lem:Order1} we know that the number of intervals stored by the pruned DPA is always smaller than 2 times the number of candidate change-points. 
Thus for $\Kmax=1$, for every $t$ $\leq n$ the pruned DPA updates on average $\mathcal{O}(\log(t))$ functional costs and intervals. From this the complexity follows $\blacksquare$

\subsection{Empirical complexity of the pDPA}\label{subsection:empiricalcomplexity}

In this section, we empirically assess the efficiency of the pDPA to analyze both simulated and real data in the case of the quadratic loss, $\gamma(Y_i, \mu) = (Y_i-\mu)^2$. The pDPA was implemented in C++ and was run on a Intel Core i7-3687U3 2.10 GHz. The code is available in the cghseg package on the CRAN at the following webpage (http://cran.r-project.org/web/packages/cghseg/index.html). 
Here is an example code using this function:
\begin{verbatim}
install.packages("cghseg")
library(cghseg)
Kmax <- 40; n <- 10^5;
y <- rnorm(n)
system.time(res_ <- cghseg:::segmeanCO(y, Kmax))
\end{verbatim}

\subsubsection{Synthetic data}
We simulated a number of sequences with a constant, a sinusoid or a rectangular wave signal.
We considered an additional gaussian noise of variance 1, a uniform noise and a Cauchy noise.
We also considered the worst case scenario for the pDPA which is, 
as explained at the end of section \ref{subsection:worst}, achieved for $y_i =i.$
We compared the pDPA \modif{}{with} the segment neighbourhood algorithm 
for $n \leq 81\ 920$. 
For all these tests, we set $\Kmax=40$. 
We then ran the pDPA alone for larger sequences $n \leq 2\ 621\ 440$.
The R code to ran the pDPA on these simulations is given in appendix \ref{app:code}.

Figures \ref{fig:DynProgComp}-A and B show that for all simulated sequences (coloured dotted and dashed lines) except the worst case scenario (black dashed and crossed line) the pDPA  is faster than segment neighbourhood (black solid line). 
It took roughly 150 seconds for the segment neighbourhood algorithm to process a sequence of $81\ 920$ data-points.
In the worst case scenario the pDPA \modif{}{was} able to process $40\ 960$ %
data-points 
in the same amount of time,
and for other simulated signals the pDPA \modif{}{was} able to process more than $2.6$ million data-points in the same amount of time.
The runtime of the pDPA depends on the nature of the noise, for example it is faster for a Cauchy noise (green dotted lines in figure \ref{fig:DynProgComp}-A, B \modif{}{and green box-plots in figure \ref{fig:DynProgComp}-C}).

\subsubsection{Real data}
We download the publicly available $GSE17359$ project from Gene Expression Omnibus (GEO: {http://www.ncbi.nlm.nih.gov/geo/}). 
This data set is made of SNP (Single Nucleotide Polymorphism) array experiments. 
SNP arrays enable the study of DNA copy number gains and losses along the genome. 
For this kind of sequence, a multiple change-point model based on the quadratic loss 
is often used \citep{picard2005statistical}. This model has been shown to 
reach state of the art performances \citep{lai2005comparative, hocking2013learning}.

\begin{figure}[H]
\begin{center}
\begin{tabular}{m{0cm}m{7.5cm}m{0cm}m{7.5cm}}
A & \includegraphics[height=6.5cm]{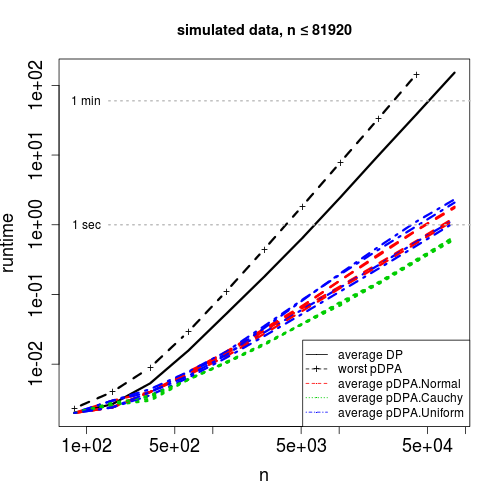} &
B & \includegraphics[height=6.5cm]{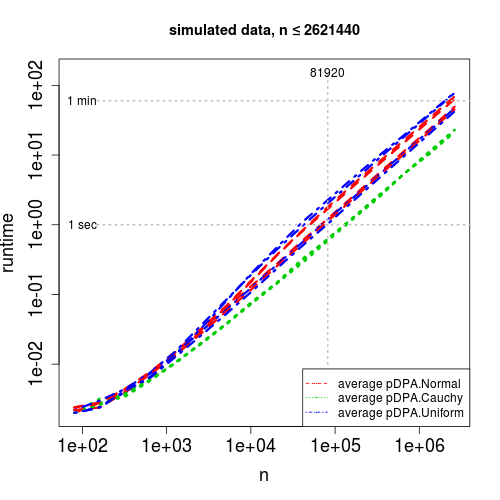} \\
C & \includegraphics[height=6.5cm]{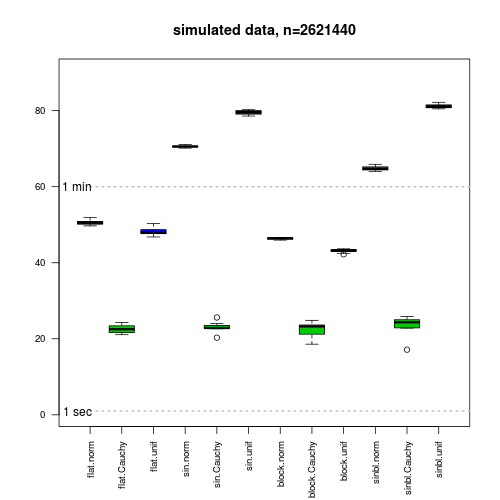}  & 
D & \includegraphics[height=6.5cm]{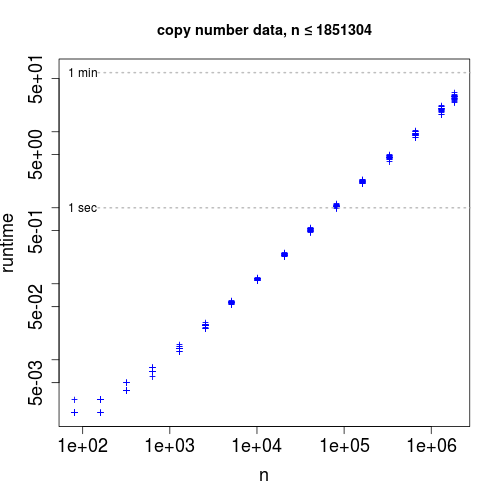} 
\end{tabular}

\end{center}
\caption{Runtimes as a function of $n$ in log-scale. 
A: Mean runtimes in seconds of the segment neighbourhood algorithm (black solid line) and pDPA (dotted, dashed, and dashed dotted line) 
for sequences of less than $81\ 920$ data-points. 
The black dashed and crossed line is the runtime of the pDPA in the worst case scenario.
Coloured dashed and dotted lines correspond to sequences simulated with or without sine or block waves plus an additional normal (red), Cauchy (green) or uniform noise (blue) (see appendix \ref{app:code}).
B: Same as A for sequences of less than $2.6$ million data-points.
\modif{}{C: Boxplot of runtimes (in seconds) of the pDPA to process simulated sequences of size $n=2621440$.}
\modif{}{D}: Runtimes (in seconds) of the pDPA to process the sequences of the $GSE17359$ dataset from data-point $1$ to $t$. 
}
\label{fig:DynProgComp}
\end{figure}

Each SNP array experiment can be viewed as a sequence of $1.8$ million data-points. We assessed the runtime
of the pDPA to process these sequences from data-point $1$ to $t$ for various value\modif{}{s} of $t$ and again with $\Kmax=40$. 
The R code is given in appendix \ref{app:code_realdata}. Runtimes are reported in  Figure \ref{fig:DynProgComp}-\modif{}{D}.
The runtime to process the $1.8$ million data-points was on average 28 seconds and was always smaller than 33 seconds.
Note that the runtime performances of the pDPA on such SNP array datasets have been confirmed by \citet{hocking2013learning}.

\begin{figure}[H]
\begin{center}
\begin{tabular}{m{0cm}m{6cm}m{0cm}m{6cm}}
A & \includegraphics[height=4cm]{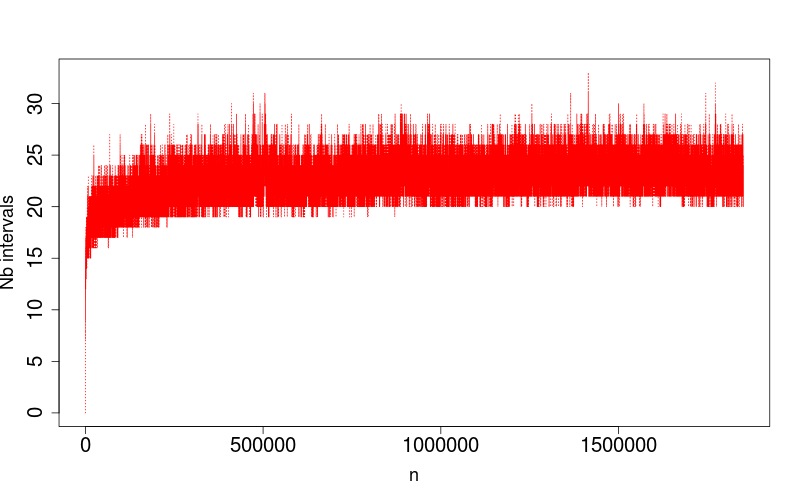} & B & \includegraphics[height=4cm]{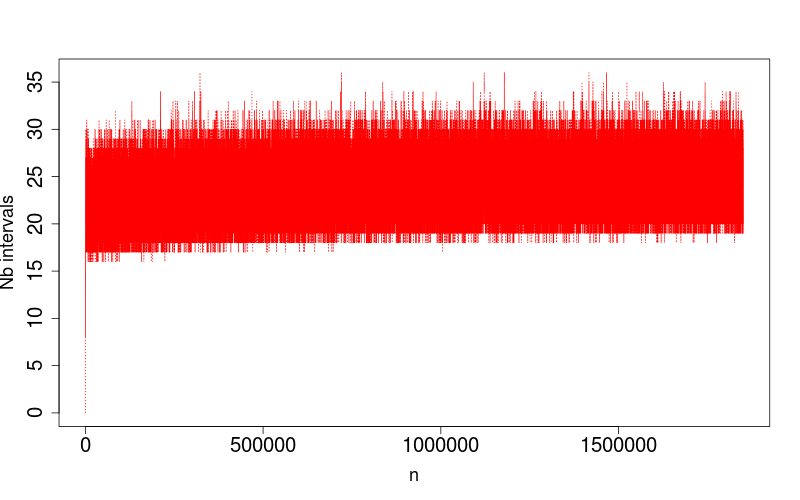} \\
C & \includegraphics[height=4cm]{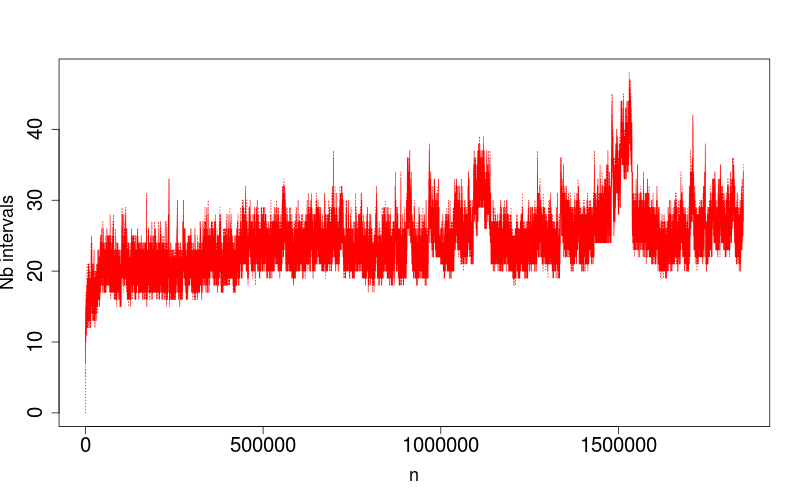} 
\end{tabular}
\end{center}
\caption{Maximum number of intervals stored by the pDPA at each point of the sequence
for $K=1$.
A: For 100 sequences of $1.8 \ 10^6$ points simulated with a constant signal plus an additional normal noise of variance 1. B:
For 100 sequences of $1.8 \ 10^6$ points simulated with a sine wave signal plus an additional normal noise of variance 1. C:
For the 18 profiles of length $1.8 .\ 10^6$ of the $GSE17359$ dataset.}
\label{fig:numberofintervals}
\end{figure}

\paragraph{Number of intervals}
We also directly assess the efficiency of the pruning by counting the number of intervals stored by the pDPA at every time step of the algorithm.
Figure \ref{fig:numberofintervals}-A, B and C show, for sequences with a constant or 
sine wave signal plus an additional gaussian noise and for sequences of the $GSE17359$ project,
 the limited number of intervals stored by the pDPA. 
Indeed, for all these sequences we observed less than 50 intervals. 
If there was no pruning we would expect at least $n = 1.8$ million and at worst $2n - 1$ intervals (see subsection \ref{subsection:worst}). 

\section{Conclusion}
This paper presented the pDPA that recovers exactly the best segmentations with $1$ to $\Kmax$ change-points of an $n$ point sequence w.r.t a loss function. 
This algorithm is based on a functional representation of the cost of segmentations.
For loss functions with a unidimensional parameter per segment the worst case complexity is in $\mathcal{O}(\Kmax n^2)$, thus the pDPA is at worst equivalent to segment neighbourhood algorithms.
For a special loss function it is proven that the pDPA allows to prune the set of candidate change-points efficiently even when there are no change-points in the signal.
Finally, in the case of the quadratic loss, it was shown empirically that the pDPA has a small runtime
compared to segment neighbourhood algorithms.

\paragraph{Future work}

Here we theoretically proved and empirically assessed that functional pruning is efficient under a restrictive set of assumptions. 
It would be interesting to more generally characterise what are the conditions for functional pruning to be efficient.
In particular functional pruning is theoretically possible for a $p$ dimensional parameter.
Recently,
\citet{maidstone2014optimal}
proved that functional pruning prunes at least as much as inequality based pruning. 
Based on this, it follows that if the conditions of Theorem 3.2 of
\citet{killick2012optimal}
are met then functional pruning with optimal partitioning (Fpop) is efficient.
However whether functional pruning is on average efficient when $p$ is large and the number of change-points is fixed 
is to our knowledge an open question.

Another question is how to implement functional pruning
in the multidimensional case. Indeed this would require an efficient way 
to represent and update the set of parameter values for which a particular change is optimal.
To the best of our knowledge these sets do not have particularly simple properties and are not necessarily easy to handle, 
except
maybe for some special loss functions such as the $\ell_{\infty}$ loss.

\paragraph{Acknowledgement}
I would like to thank Alice Cleynen, Michel Koskas, Robert Maidstone, Toby Hocking, Paul Fearnhead, Emilie Lebarbier, Stéphane Robin for fruitful discussions that helped me to better
understand the pDPA. Special thanks to Toby, Alban and Patrick.

\bibliography{biblio_new}

\appendix
\newtheorem{de}{Definition}[subsection]

\section{Bound on the number of intervals}\label{Section:LemmaProof}
In this section, we study a special class of functions that we denote $\ConvBunch$ and demonstrate theorem \ref{Lem:Order1}. A direct application of this theorem 
shows that if there are $t$ candidate last change-points, 
then the pDPA need to store at most $2t-1$ intervals. 
We used this theorem in section \ref{section:algo} 
to prove the worst-case complexity of the pDPA.

\subsection{ $\ConvBunch$ functions}
\begin{de}
\label{Def:HConc}
Let $\ConvBunch_n$ denote the set of all functions $B: \mathbb{R} \rightarrow \mathbb{R}$ such that
\begin{eqnarray*}
\forall \ \mu \ \in \ \mathbb{R}, \ B(\mu) = min_{\{ t \ \in \ \UnN \} } \ \{ u_{B,t} + \sum_{j = t+1}^{n+1} f_{B,j}(\mu) \}
\end{eqnarray*}
where all $u_{B,t}$ are real numbers and all $f_{B,j}$ are 

unimodal functions of $\mu$ \footnote{Here by unimodal we mean functions such that for any $c$ in $\mathbb{R}$ the set $\{ x | f(x) \leq c\}$ is an interval.}  
and any $\sum_j f_{B,j}(\mu)$ is also unimodal . Note that $\ConvBunch_n \subset \ConvBunch_{n+1}$.
Let $\ConvBunch$ $= \bigcup \ConvBunch_n$. 

\end{de}
If the loss function $\gamma(y_i, \mu)$ and any $\sum_i \gamma(y_i, \mu)$ are unimodal in $\mu$, 
$\FCost_{\OnetoT}^K(\mu)$ is part of $\ConvBunch_{t-1}$ because we have: $\FCost_{\OnetoT}^K(\mu) = \underset{\tau < t}{\min} \{ \FCc_{\OnetoT, \tau}^K(\mu) \}$.
\begin{de}
For any $B$ $\in$ $\ConvBunch_{n}$ and $A$ a subset of $\ \UnN$ we define the function $B_A$ as
$$\forall \ \mu, \ B_A(\mu) = min_{\{ t \ \in \ A \} } \ \{ u_{B,t} + \sum_{j = t+1}^{n+1} f_{B,j}(\mu) \}$$
\end{de}
\begin{prop}
$B_A \in \ConvBunch_{card(A)}$ 
\end{prop}
It is easily shown for $A= \UnN \setminus \{i \}$ with $i$ $\in$ $\ \UnN$ and thus by induction it is true for any $A$.
\begin{de}
The rank of a function $ B \in \ConvBunch$ is $\rank(B) = min \{ n \ \in \mathbb{N}^* \ | B \in \ConvBunch_n \} $
\end{de}
\subsection{Decomposition in intervals and order of $\ConvBunch$}
\begin{de}
Let $\mathcal{I}$ be a partition of $\mathbb{R}$ in a finite set of intervals 
$\mathcal{I} = \{I_j\}_{ \{j \in \UnK \}} $. $\mathcal{I}$
is a $k$-decomposition of a function $B \ \in \ \ConvBunch$ if 
\begin{eqnarray*}
\forall \ I_j, \ \exists \ i, \ \forall x \ \in \ I_j \qquad B_{{i}}(x) = B(x)
\end{eqnarray*}
The set of all $\ConvBunch$ functions with a $k$-decomposition is denoted $\ConvBunch^k$.
Similarly the set of all $\ConvBunch_n$ functions with a $k$-decomposition is denoted $\ConvBunch_n^k$.
\end{de}

\begin{de}
The order $O(B)$ of a $\ConvBunch$ function is
$ min \{ k \in \mathbb{N}^* | B \ \in \ \ConvBunch^k \ \}$
\end{de}
\begin{theo}
\label{Lem:Order1}
For all $B \in \ConvBunch$, we have $\order(B) \leq 2 \times \rank(B) - 1$.
\end{theo}
\paragraph{Proof}
We demonstrate this theorem by induction. It is true if $\rank(B) \leq 1$. Assume it is true for any $B$ with $\rank(B) \leq n$.
Let $B$ $\in$ $\ConvBunch$ with $\rank(B)=n+1$. 
We have:
\begin{eqnarray*}
\forall \mu \in \mathbb{R}, \qquad & B(\mu) & = min \{ B_{ \UnN }(\mu),\ B_{\{ n+1 \}}(\mu) \} \\
& B(\mu) & = min \{ C(\mu),\ u_{B, n+1} \} + f_{B,n+2}(\mu),
\end{eqnarray*} where $C \in \ConcBunch_n$:
$$ C(\mu) = \min_t \{ u_{B,t} + \sum_{j = t+1}^{n+1} f_{B,j}(\mu) \}$$
Let $\mathcal{I} = \bigcup_{j \in \UnK} {I_j}$ 
be the smallest set of intervals such that: $$ \forall \ I_j \ \in \ \mathcal{I} , 
\ \forall \ \mu \ \in \ I_j \qquad u_{B, n+1} > \ C(\mu)$$
Let $A_k$ be the subset of $\UnN$ defined as 
$\{ i \quad | \quad \exists \ x \ \in \ I_k \quad C_{\{ i \}}(x) < u_{B, n+1} \}$.

As $\rank(B)=n+1$ and as for all $i$ in $\UnN$ $C_{\{ i \}}$ is unimodal, there exists a unique
$j$ such that $i \in A_j$ and therefore $\sum_{j=1}^k card(A_j) = n$.
In each interval $I_k$, we have $C(\mu) = C_{A_k}(\mu)$. By induction, $\order(C_{ A_k }) \leq 2 \times card(A_k) - 1$.
Overall, for any $B$ with $\rank(B)=n+1$, we have:
\begin{eqnarray*}
\order(B) \leq & \sum_{j=1}^k \order(C_{A_k }) + (k+1) \leq & 2 \sum_{j=1}^k card(A_k) +1 \leq 2n+1 \leq 2\rank(B)-1 \quad \blacksquare
\end{eqnarray*}

\section{Lower bound on the probability that $\Sc_{\OnetoN, \tau}^1$ is empty }\label{app:averagecomp}
\paragraph{Model}
We consider the negative log-likelihood loss of a continuous uniform distribution defined on $[0, \mu]$.
The loss function is:
\begin{eqnarray*}
\gamma(y_i, \mu) = & \log(\mu) &\mbox{if} \ \mu \geq Y_i \\
\gamma(y_i, \mu) = & \infty & \mbox{otherwise}
\end{eqnarray*}
This loss function is unimodal and any sum of $\gamma(Y_i, \mu)$ is also unimodal. 
Indeed, for any $t \geq \tau \geq 1$ we have:
\begin{eqnarray*}
\sum_{\tau +1}^t \gamma(Y_i, \mu) = & (t- \tau)\log(\mu) & \mbox{if} \ \mu \geq \hat{y}_{\tau,t}\\
\sum_{\tau +1}^t \gamma(Y_i, \mu) = & \infty & \mbox{otherwise}
\end{eqnarray*}
We define $\hat{y}_{\tau:t} = \max_{i \in \{ \tau \ldots t-1 \} } \{ y_i \}.$
We have 
$$\Cost_{\OnetoT}^0 = \min_{\mu \in \mathbb{R}^+_*} \{ \sum_{1}^{t-1} \gamma(Y_i, \mu)\} = 
(t-1) \log(\hat{y}_{1:t}).$$
Finally note that the loss is continuous to the right and as explain\modif{}{ed} in the footnote page \pageref{footnote:1} we can further restrict 
$\Tauc_{\OnetoT}^{K}$ to the set of $\tau$ such that $\Sc_{\OnetoT, \tau}$ is not restricted to a single value.

\paragraph{Proof}
Using equation \ref{eq:set2} we get that for any $t$ greater than 2 we have:
\begin{eqnarray}
\Sc_{\OnetoT+1, t}^1 & \subset & \{ \mu \ | \ \Cost_{\OnetoT}^{0} \leq  \FCc_{\OnetoT, t}^1(\mu) \} \\
& = &]0, y_{t-1}] \cup [\hat{y}_{1:t}. \left(\frac{\hat{y}_{1:t}}{\hat{y}_{1:t-1}}\right)^{t-1}, +\infty[, \\
\end{eqnarray}
with $\left(\frac{\hat{y}_{1:t}}{\hat{y}_{1:t-1}}\right)^{t-1} \geq 1.$
There are two cases in which it can be shown that $\Sc_{\OnetoN, t}^1$ 
is empty or restricted to a single value and thus can be discarded.

\paragraph{Case 1}: If $ y_{t-1} < y_{t} < \hat{y}_{1:t} $ we have that 
$\hat{y}_{1:t+1} = \hat{y}_{1:t} = \hat{y}_{1:t-1}$. Thus we have

\begin{eqnarray*}
\Sc_{\OnetoT+1, t}^1 & \subset & ]0, y_{t-1}] \cup [\hat{y}_{1:t}, +\infty[, \\
\{ \mu \ | \ \FCc_{\OnetoT+1, t}^1(\mu)  & \leq & \Cost_{\OnetoT+1}^{0} \} = [y_{t}, \hat{y}_{1:t+1}].
\end{eqnarray*}
and using equation \ref{eq:set1} from property \ref{prop:set} that:

$$\Sc_{\OnetoT+2, t}^1 \subset \left(]0, y_{t-1}] \cup [\hat{y}_{1:t}, +\infty[\right)  \cap [y_{t}, \hat{y}_{1:t}] = \{y_{t}\}$$
and by induction we get $\Sc_{\OnetoN, t}^1 \subset \{y_{t}\}$.

\paragraph{Case 2}: If $y_{t} < y_{t-1} < \hat{y}_{t:n}$ we have $\hat{y}_{1:t+1} = \hat{y}_{1:t}$ and we get using equation \ref{eq:set1}:
\begin{eqnarray*}
\Sc_{\OnetoT+2, t}^1 \subset & \left(]0, y_{t-1}] \cup [\hat{y}_{1:t}. \left(\frac{\hat{y}_{1:t}}{\hat{y}_{1:t-1}}\right)^{t-1}, +\infty[ \right) \quad \bigcap \quad [y_{t}, \left(\hat{y}_{1:t+1}.\frac{\hat{y}_{1:t+1}}{\hat{y}_{1:t}}\right)^{t}] \\
\subset & [y_t, y_{t-1}].
\end{eqnarray*}
Now we also have:
\begin{eqnarray*}
\{ \mu \ | \ \FCc_{\OnetoN,t}^{1}(\mu) & \leq & \Cost_{\OnetoN}^0 \} =
\left[\hat{y}_{t:n}, \left(\hat{y}_{1:n}.\frac{\hat{y}_{1:n}}{\hat{y}_{1:t}}\right)^{\frac{t-1}{n-t}}\right]
\end{eqnarray*}
Using equation \ref{eq:set1} we get that $\Sc_{\OnetoN, t}^1 \subset [y_t, y_{t-1}] \cap \{ \mu \ | \ \FCc_{\OnetoN,t}^{1}(\mu)  \leq  \Cost_{\OnetoN}^0 \} = \emptyset,$ as $y_t < \hat{y}_{t:n}.$

\paragraph{}
Let us now compute the probability of these two events.

All $Y_i$ are independent, they have the same density $f$ and continuous distribution $F$, thus the distribution of $\hat{Y}_{1:t}$ is $F^{t-1}$ and $P(\hat{Y}_{1:t} < x) = P(\hat{Y}_{1, t} \leq x)=F^{t-1}(x)$.
As $Y_{t}$ is independent of $\hat{Y}_{1:t}$ we get that 
$P(\hat{Y}_{1:t} < Y_{t:t+1} ) = \int_\mathbb{R} \! 2 f(x) F(x) P(\hat{Y}_{1:t} < x)\mathrm{d}x = \int_\mathbb{R} \! 2f(x)F(x)^t\mathrm{d}x = \frac{2}{t+1}.$

Using this, we see that case 1 happens with probability: 
$$P(Y_{t-1} < Y_{t} < \hat{Y}_{1:t}) =P(Y_{t-1} \leq Y_{t} \leq \hat{Y}_{1:t-1})= \frac{1}{2}(1 - \frac{2}{t}).$$ 
and case 2 happens with probability:
$$P(Y_{t} < Y_{t-1} < Y_{t:n}) = P(Y_{t} \leq Y_{t-1} \leq Y_{t+1:n}) =\frac{1}{2} (1 - \frac{2}{n-t}).$$
Furthermore case 1 and case 2 are disjoint events. Thus the probability
that the two occur is the sum $1 - \frac{2}{t} - \frac{2}{n-t}$.
Thus the probability that the change-point $t$ has not been discarded at step $n$
is lower than $\frac{2}{t} + \frac{2}{n-t}$. 
So taking the \modif{}{expectation} and summing \modif{}{across} all $t$ smaller than $n$\modif{}{, we} recover that the expected number of non discarded candidate change-points is in $\Oc(\log(n))$.
$\blacksquare$

\section{R code to assess the runtime of pDPA on simulated data}\label{app:code}
Here we provide the R code we used to assess the performances of the pDPA.
The code of the pDPA is in C++.

First here is a function to simulate different type of signals.

\begin{verbatim}
#### simple simulation function
funSimu_ <- function(n, noise="norm", signal="flat"){
	### signal
	if(signal == "flat"){
		x <- numeric(n) 	
	}
	if(signal == "sin"){## sin
		x <- 2*sin(1:n/100)	
	}
	if(signal == "block"){## 10 blocks
		nb=5
		x <- rep(rep(c(0,2), nb), each=n/(2*nb))[1:n]		
	}
	if(signal == "sinblock"){## 10 blocks + sin
		nb=5
		x <- rep(rep(c(0,2), nb), each=n/(2*nb))[1:n]	+ 2*sin(1:n/100)
	}
	if(signal == "linear"){
		x <- 1:n	
	}
	
	### noise
	if(noise == "none") ### do nothing
	if(noise == "norm") x <- x + rnorm(n)
	if(noise == "Cauchy") x <- x +rcauchy(n)
	if(noise == "unif") x <- x+ runif(n,0, 1) -0.5
	return(x)
}

\end{verbatim}

Second, here is a simple function to record the runtime of the algorithm.
\begin{verbatim}
simpleStat <- function(x){c(mean(x), sd(x), range(x))}
#### simple function to test the runtimes for various n 
testRunTime <- function(Kmax = 50, ns=c(100, 200), rep=5, 
		funSimu=rnorm, noise, signal){
	rTime <- list()
	for(n in ns){
	cat("Size:", n , ", Repetition 1/", rep,  "\n")
	for(j in 1:rep){
	x <- funSimu(n, noise=noise, signal=signal)
	rTime[[length(rTime)+1]] <- c("pDPA", n, j, 
		system.time(res_p <- cghseg:::segmeanCO(x, Kmax=Kmax))[3])
	}
	}
	### formating the runtimes
	dTime <- data.frame(matrix(unlist(rTime), ncol=4, byrow=TRUE))
	colnames(dTime) <- c("Algo", "n", "rep", "time")
	dTime$time <- as.numeric(as.vector(dTime$time))
	dTime$n <- as.numeric(as.vector(dTime$n))
	mTime <- aggregate(dTime$time, by=list(dTime$Algo, dTime$n), 
		FUN= simpleStat, simplify=TRUE )
	mTime <- cbind(mTime[, 1:2], mTime[, 3][, 1:4])
	colnames(mTime) <- c("Algo", "n", "mean", "sd", "min", "max")
	mTime$signal <- signal
	mTime$noise <- noise
	mTime <- mTime[order(mTime$n), ]
	return(mTime)	
}

\end{verbatim}

Here is a code to run the algorithm for different types of signal (constant, sinusoid, block), various types of noise
(gaussian, uniform and Cauchy) and different values of $n$.
\begin{verbatim}

install.packages("cghseg")
require(cghseg)

### 20 000 +
Kmax=40
ns <- 40* 2^(1:16)
respDPA_l <- list()


for(signal in c("flat", "sin", "block", "sinblock")){
print(paste("###", signal))
for(noise in c("norm", "Cauchy", "unif")){
print(noise)
respDPA_l[[length(respDPA_l)+1]] <- testRunTime(Kmax = Kmax, ns=ns, rep=5, 
		funSimu=funSimu_, noise=noise, signal=signal)
}}
save(respDPA_l, file="respDPA_l.Rdata", compress=TRUE)

\end{verbatim}

Here is a code to run the algorithm in the worst case scenario, $Y_i=i.$
\begin{verbatim}
Kmax=40
ns <- 40* 2^(1:10)
respDPA_w <- list()

respDPA_w <- list()
for(signal in c("linear")){
print(paste("###", signal))
for(noise in c("none")){
print(noise)
respDPA_w[[length(respDPA_w)+1]] <- testRunTime(Kmax = Kmax, ns=ns, rep=5, 
		funSimu=funSimu_, noise=noise, signal=signal)
}}
save(respDPA_w, file="respDPA_w.Rdata", compress=TRUE)
\end{verbatim}

\section{R code to assess the runtime of the pDPA on the GSE17359 data set}\label{app:code_realdata}
Here we provide the R code we used to assess the performance of the pDPA
on the data from the GSE17359 data set.
\begin{verbatim}
dat <- read.table("GSE17359_affy_snp_ratios_matrix.txt", header=TRUE, sep="\t")
require(cghseg)
ie <- which(apply(is.na(dat), 1, sum) == 0)
dat <- dat[ie,]
ns <- c(40* 2^(1:15), nrow(dat))
Kmax = 40
rTime <- list()
for(iSample in 2:ncol(dat)){
print(iSample)
	for(n in ns){
	x <- dat[1:n, iSample]
	rTime[[length(rTime)+1]] <- c("pDPA", n, iSample, 
			system.time(res_p <- cghseg:::segmeanCO(x, Kmax=Kmax))[3])
	}
}

dTime <- data.frame(matrix(unlist(rTime), ncol=4, byrow=TRUE))

\end{verbatim}

\end{document}